\def\al26{\mbox{$^{26}$\hspace{-0.2em}Al}}          
\def\MeV{\mbox{Me\hspace{-0.1em}V}}                 
\def\keV{\mbox{ke\hspace{-0.1em}V}}                 
\def\deg{\mbox{$^\circ$}}                           
\def\dg{^\circ}                                     
\def\funit{photons cm$^{-2}$ s$^{-1}$}              
\def\phibar{$\bar{\varphi}$}                        
\def\Msol{\mbox{M$_{\odot}$}}                       
\def\HII{\mbox{H\hspace{0.2em}{\scriptsize II}}}    
\def\etal{{et al.}}                                 
\def\etacar{$\eta$~Carinae}                         
\def\gray{\mbox{$\gamma$-ray}}                      
\def\reference{}
\def\AA{A\&A}                                       
\def\ApJ{ApJ}                                       

\documentstyle[psfig]{l-aa}

\begin{document}

\thesaurus{05(13.07.2; 02.14.1; 10.15.1; 08.19.4; 08.23.2)}

\title{1.8 \MeV\ Emission from the Carina Region}

\author{J.~Kn\"odlseder$^{1,5}$, K.~Bennett$^4$, H.~Bloemen$^2$,
	    R.~Diehl$^1$, W.~Hermsen$^2$, U.~Oberlack$^1$,  J.~Ryan$^3$,
	    and V.~Sch\"onfelder$^1$}

\institute{$^1$Max-Planck-Institut f\"ur extraterrestrische Physik,
	   Postfach 1603, 85740 Garching, Germany\\
	   $^2$SRON-Utrecht, Sorbonnelaan 2, 3584 CA Utrecht, The
	   Netherlands\\
	   $^3$Space Science Center, University of New Hampshire, Durham
	   NH 03824, U.S.A.\\
	   $^4$Astrophysics Division, ESTEC, ESA, 2200 AG
	   Noordwijk, The Netherlands\\
	   $^5$Centre d'Etude Spatiale des Rayonnements, CNRS/UPS, BP 4346,
	   31029 Toulouse Cedex, France}

\offprints{J\"urgen Kn\"odlseder (Toulouse)}

\date{Received October 1995; accepted October 1995}

\maketitle
\markboth{J. Kn\"odlseder et al.: 1.8 \MeV\ Emission from the Carina Region}
	 {J. Kn\"odlseder et al.: 1.8 \MeV\ Emission from the Carina Region}

\begin{abstract}
Significant 1.8 \MeV\ emission from the Carina region has been detected
by COMPTEL.
The emission is concentrated within 6 degrees or less near the Carina nebula
NGC 3372, one of the brightest \HII\ regions known in our Galaxy.
This region contains a wealth of extreme young open clusters
whose massive stars possibly contributed to an enrichment of \al26\
in the ISM within the last few million years.
The relation of these clusters and the peculiar object \etacar\ with the
observed emission is discussed.
The \al26\ yield of the clusters is estimated using current theoretical
nucleosynthesis models.

\keywords{gamma rays: observations -- nucleosynthesis -- open clusters --
          supernovae -- stars: Wolf-Rayet}

\end{abstract}

\section{Introduction}

\al26\ is the first radioactive isotope which was detected by its
\gray\ line emission in the interstellar medium (\cite{rf:mahoney}).
With its short lifetime of $\sim~10^6$ yrs, it is
a clear tracer of ongoing nucleosynthesis in our Galaxy.
\al26\ is assumed to be produced in various sites, such as
core-collapse supernovae (SNe), Wolf-Rayet (WR) stars, O-Ne-Mg novae,
and asymptotic giant-branch (AGB) stars (see review of \cite{rf:pd95}).
After two years of observations, the imaging telescope COMPTEL aboard
the CGRO satellite revealed the first map of the Milky Way in the
light of the \al26\ decay line at 1.809 \MeV\ (\cite{rf:diehl95a}).
The sky map clearly shows that the emission is confined to the Galactic
plane.
Besides the important emission from the central Galactic radian,
1.8 \MeV\ emission is seen in the direction of Cygnus, Vela, and
near the anticentre.
One of the most prominent features is found at $(l,b)=(286.5\dg,0.5\dg)$
in the direction of Carina.
Among all features in the map it appears to be the most concentrated
one, and additionally it lies in a nearly emission free region of the plane.
We present here a detailed study of this region and discuss the
potential source candidates.

\section{Data Analysis and Results}

COMPTEL allows the study of 1.8 \MeV\ \gray\ line emission with an energy
resolution of $\sim8\%$ (FWHM) and an angular resolution of 1.6\deg\
($1\sigma$) within a wide field of view of about 1 steradian.
For an observation time of $10^6$ seconds a \gray\ line sensitivity of
some $10^{-5}$ \funit\ is obtained.
Incoming \gray\ photons are measured by their consecutive interactions in
two parallel detector planes where an incident photon is first
Compton scattered in the upper layer and then absorbed (although often
not completely) in the lower layer.
{}From the energy deposits and the interaction location in both layers,
the Compton scatter angle \phibar\ and the scatter direction ($\chi,\psi$)
are calculated, which span the three-dimensional imaging data-space.
A detailed description of the instrument can be
found in Sch\"onfelder \etal\ (1993).
For the 1.8 \MeV\ \gray\ line analysis, only events in a 200 \keV\ wide
energy window centred on 1.8 \MeV\ are used.
The majority of the registered photons ($\ga95\%$) are due to instrumental
background.
We estimated their distribution in the data-space by measurements at
adjacent energy intervals which we corrected for the energy dependence of
the Compton scatter angle \phibar\ (\cite{rf:knoedl95a}).
Data-space analysis in this approach suppresses the continuum emission
and reveals only pure 1.8 \MeV\ line emission.
We used data from CGRO observation phases I+II (May 1991 - August 1993) for
our analysis.
During this period, the Carina region was in the field of view for
$\sim120$ days, although half of the time at rather large aspect angles.

\begin{figure}
 \setlength{\unitlength}{1cm}
 \begin{minipage}[t]{8.8cm}
  \begin{picture}(8.8,5)
   \framebox(8.8,5){
    \psfig{file=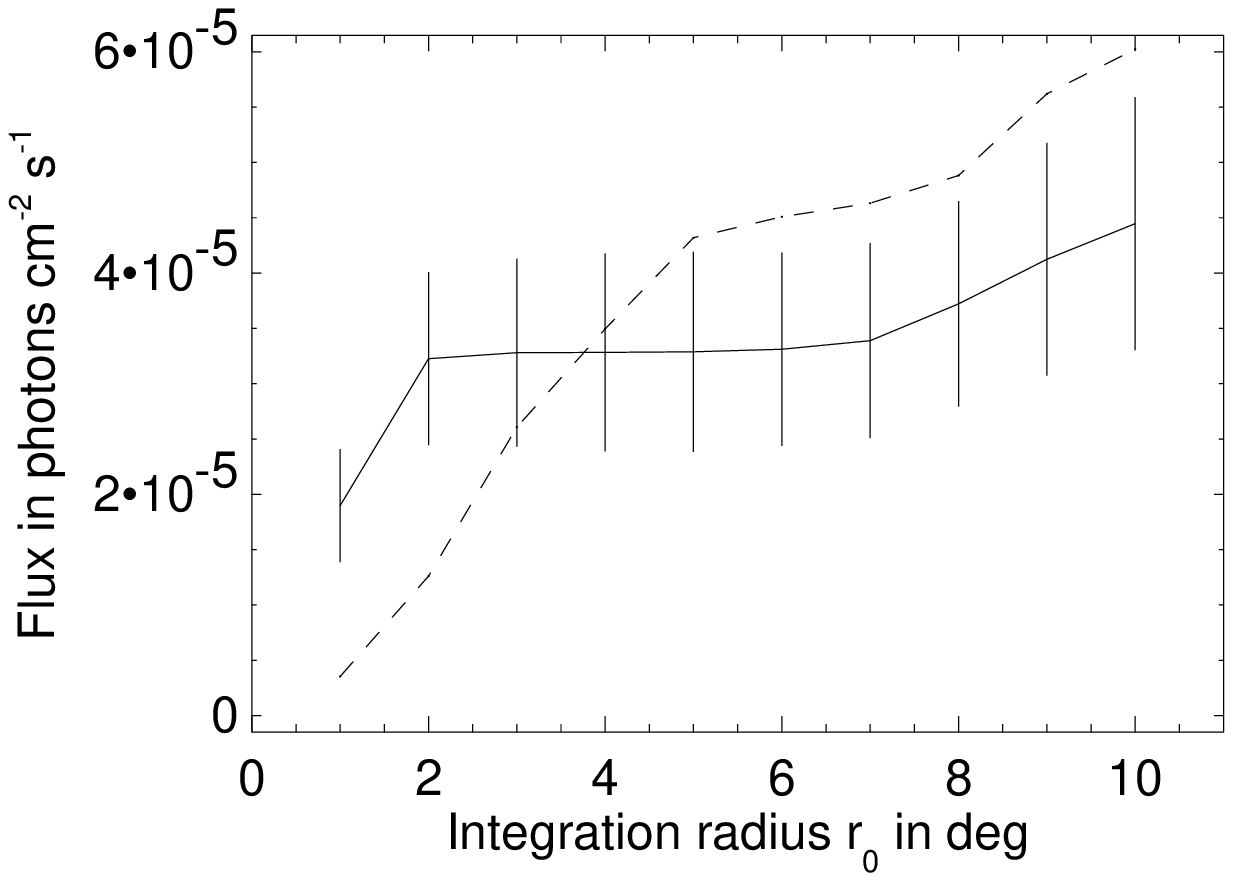,height=5cm}}
  \end{picture}
  \caption{\label{fig:flux}
    Growth curve derived from the maximum entropy intensity map
    for the point-like 1.8 \MeV\ feature in Carina (solid) and
    the more extended emission in Vela (dashed; Diehl \etal\ 1995b).}
 \end{minipage}
\end{figure}

\begin{figure}
 \setlength{\unitlength}{1cm}
 \begin{minipage}[t]{8.8cm}
  \begin{picture}(8.8,8.8)
   \put(0,0){\framebox(8.8,8.8){
    \psfig{file=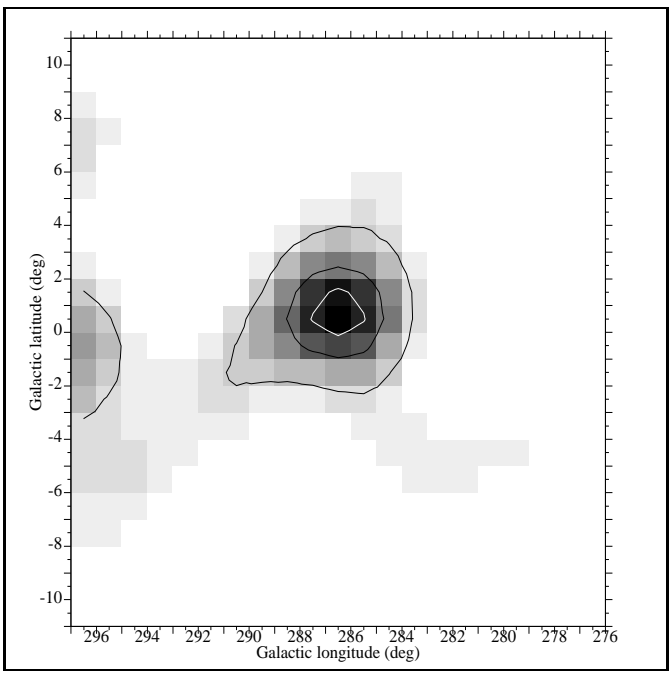,width=8.8cm,height=8.8cm}}}
  \end{picture}
  \caption{\label{fig:carreg}
    COMPTEL 1.8 \MeV\ maximum likelihood map of the Carina region.
    The contour lines indicate the $1\sigma$, $2\sigma$,
    and $3\sigma$ error regions of the source location.}
 \end{minipage}
\end{figure}

Two different techniques were used to derive the 1.8 \MeV\ flux and
the extension of the emission in Carina.
First, we applied the maximum entropy method (\cite{rf:strong92})
to reconstruct a deconvolved 1.8 \MeV\ intensity map of the Carina region.
{}From integration of the sky map over circular areas with increasing radii,
centered on the emission maximum at $(l,b)=(286.5\dg,0.5\dg)$,
we obtained a `growth curve' of the 1.8 \MeV\ emission
(see Fig. \ref{fig:flux}).
It shows that most of the emission lies within a radius of $r_0\approx2\dg$
around the maximum which is consistent with the appearance of a
point source.
The flatness of the growth curve between 2\deg\ and 7\deg\ demonstrates
that the emission feature is isolated in a region free of 1.8 \MeV\
emission.
{}From the plateau in the growth curve we infer a total flux of
$(3.3\pm0.8)\times10^{-5}$ \funit\ for the entire emission feature.
The statistical flux error was derived by means of a bootstrap analysis
(\cite{rf:knoedl95a}).

Secondly, we used the maximum likelihood technique (\cite{rf:deBoer92})
to test for the statistical significance of the 1.8 \MeV\ detection
and to constrain the true emission diameter.
In this approach, a point-source or an alternate model is convolved
into the data-space which is then fitted together with the instrumental
background model to the data.
We scanned the Carina region with a point-source hypothesis and
present the result as a significance sky map in Fig. \ref{fig:carreg}.
The highest evidence for 1.8 \MeV\ emission is found at
$(l,b)=(286.5\dg,0.5\dg)$ with a detection significance of $4\sigma$
over background.
We obtained from the fit a 1.8 \MeV\ flux of $(3.1\pm0.8)\times10^{-5}$
\funit\
which is consistent with the maximum entropy result.
We estimated the extension of the emission by a fit of extended
symmetric source models of different radii to the data.
The best fit was found for a point source hypothesis.
{}From the decrease of the likelihood with increasing source radius, we
derive an upper limit of 5.6\deg\ ($2\sigma$) for the true diameter of the
1.8 \MeV\ emission.

\section{Discussion}

\subsection{The source type}

\begin{figure*}
 \setlength{\unitlength}{1cm}
 \begin{minipage}[t]{18cm}
  \begin{picture}(18,7.23)
   \put(0,3.90){\makebox(18,3.33){
    \psfig{file=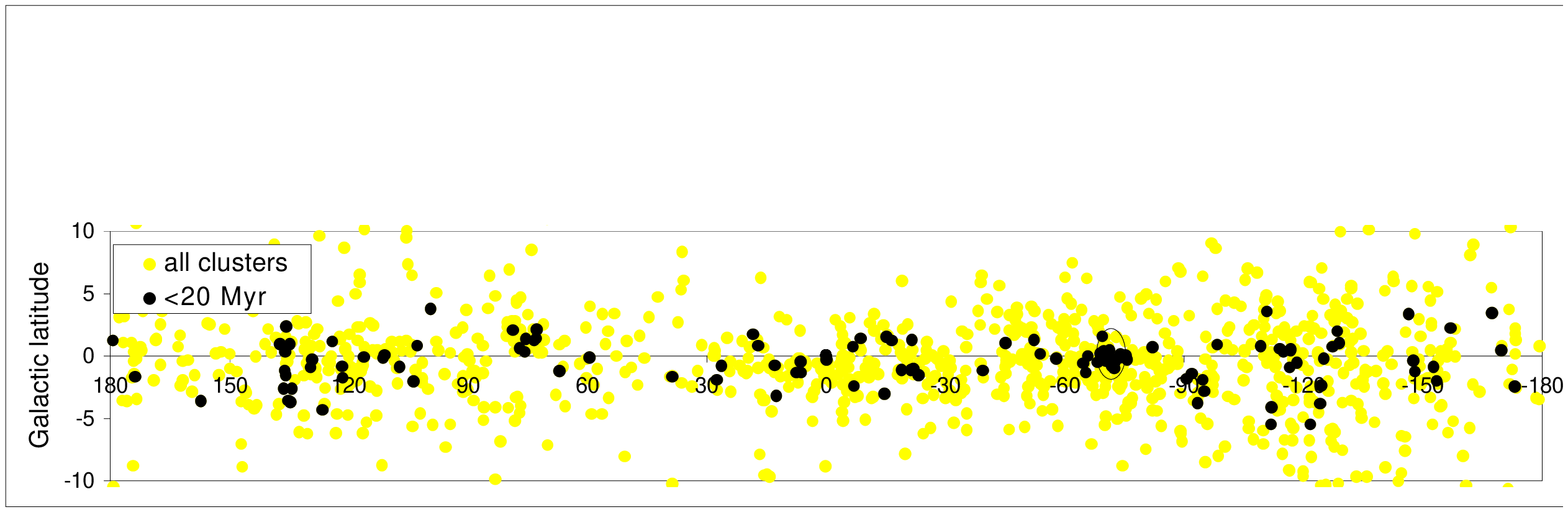,width=18cm,height=3.33cm,angle=270,clip=}}}
   \put(0,0){\makebox(18,3.9){
    \psfig{file=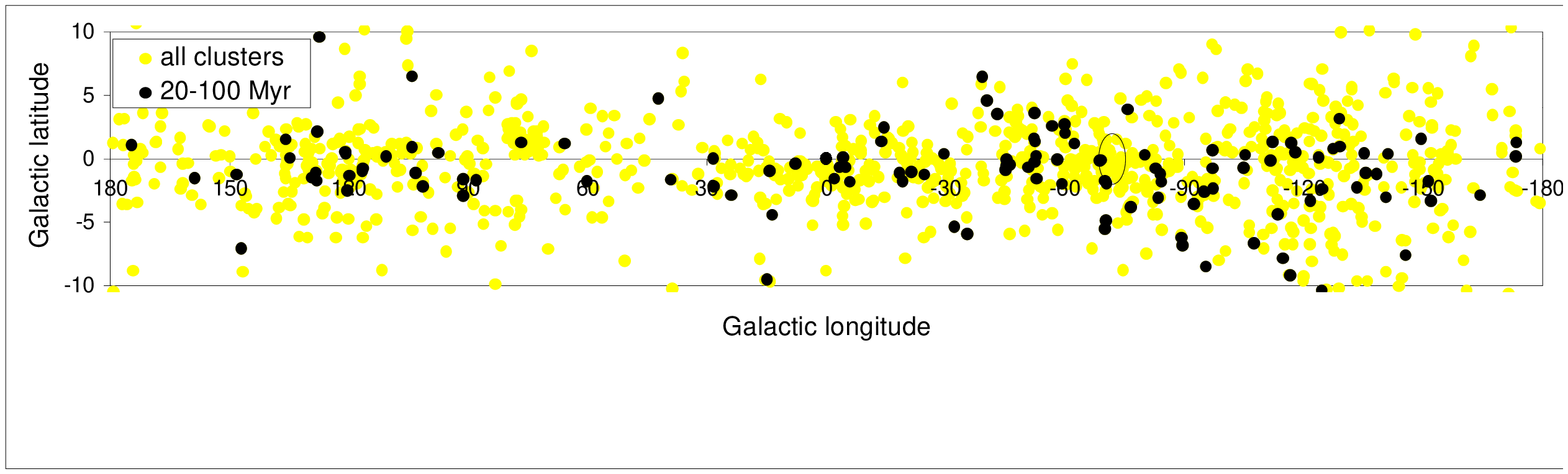,width=18cm,height=3.9cm,angle=270,clip=}}}
   \put(0,0){\framebox(18,7.23){}}
  \end{picture}
  \caption{\label{fig:lynga}
     Distribution of open clusters in the Galactic plane. The dark
     circles in the upper panel mark clusters with ages of less than 20
     Myr while in the lower panel, cluster in the age range 20-100 Myr
     are shown. Older clusters or clusters without age information are
     marked by light circles. The measured emission feature is
     indicated as ellipse.}
 \end{minipage}
\end{figure*}

The point-like appearance of the 1.8 \MeV\ emission in Carina
restricts the possibilities for its origin.
If the feature is produced by a single object, it must be rather nearby
because the \al26\ yield which is expected for individual sources is
relatively low.
Assuming a rather optimistic nova yield for \al26\ of $10^{-6}$ \Msol\
as determined recently for the most promising subtype of O-Ne-Mg novae
(\cite{rf:coc95}) one obtains a maximum nova distance of 20~pc for the
measured flux of $3.3\times10^{-5}$ \funit.
An AGB star with \al26\ yield in the range of $10^{-9}$ to $10^{-5}$
\Msol\ (\cite{rf:bazan93}) should be not at distances above 60~pc, and
also a massive WR star or SN with \al26\ yields up to $10^{-3}$ \Msol\
(\cite{rf:langer95}, \cite{rf:hoffman95}) must be closer than 600~pc.
Additionally, all potential sources eject \al26\ at rather
high velocities in the ISM, either by stellar winds or explosions.
This implies that bubbles or shells of ejected matter build up
around these objects with radii in the order of 10 pc
(\cite{rf:castor75}).
Accordingly, the 1.8 \MeV\ emission should also be extended
for such nearby objects, and the radius of the observed emission severely
constrains the age of the \al26\ ejection.

Within the error region of the 1.8 \MeV\ emission feature, we found no
indication of an \al26\ candidate source which fulfills the above
constraints.
Thus, the source in Carina is probably not a single object but rather
a sum of more objects.
It is hard to explain how novae and low- or intermediate-mass AGB stars
could accumulate so that they produce the observed feature.
Their \al26\ yield is relatively small which would require a large number
of them within a small Galactic area to produce the observed 1.8 \MeV\
emission.
Their evolution time from birth until \al26\ ejection, however, is longer
than the Galactic revolution period of $\sim10^8$~yrs which leads to a
dispersion of these objects in the Galactic disk and hence to a rather
smooth emission profile.

High-mass stars, on the other hand, are known to be produced in groups
which are visible as OB-associations and open clusters (\cite{rf:garmany94}).
The short massive-star lifetime of $10^6-10^7$~yrs (\cite{rf:schaller92})
assures that the \al26\ is ejected soon after birth of
the star and thus remains confined to the cluster region.
Typical diameters of OB-associations are around 100~pc
(\cite{rf:garmany94}).
Thus, if the observed 1.8 \MeV\ emission in Carina originates from such an
association, the upper limit for the emission diameter of 5.6\deg\ implies
a lower limit for the association distance of 1~kpc.
The \al26\ yield of massive stars is relatively high, hence only a moderate
number of individual source objects is needed.
Assuming a typical WR star or SNe yield of $10^{-4}$ \Msol\
(\cite{rf:langer95}, \cite{rf:hoffman95}) and an association distance of
1~kpc, the measured 1.8 \MeV\ flux requires \al26\ ejection from only 30
(average) massive stars within the last million years.
Therefore we feel that massive stars which are confined to stellar clusters
are the most probable source of the 1.8 \MeV\ emission in Carina.
Using the stellar lifetimes of Schaller \etal\ (1992) one finds that
in clusters younger than $\sim20$~Myr, only WR stars and SNe
should contribute significantly to an \al26\ enrichment,
while in clusters with ages between 20 to 100 Myr the massive AGB stars
should be the dominant sources of \al26.

\subsection{Young open clusters in Carina}

In order to substantiate these considerations, we investigated the open
cluster content of the Carina region.
The most comprehensive data-base of open clusters was compiled by Lyng{\aa}
(1987).
For clusters younger than 20/100 Myr it is found to be complete up to
distances of 2.3/1.6 kpc, respectively (\cite{rf:jtl88}).
In Fig. \ref{fig:lynga} we show the distribution of the clusters along
the Galactic plane, separated in clusters younger than 20 Myr (upper panel)
and clusters with age estimates of 20-100 Myr (lower panel).
The largest concentration of clusters younger than 20 Myr is found in
the direction of Carina near $l=287\dg$ within a radius of about 3\deg.
Its position and extension is consistent with the observed 1.8 \MeV\
feature.
There is no other region with similar high cluster content, neither in
the 0-20 Myr nor in the 20-100 Myr age range.
We found 31 clusters in Lyng{\aa}'s catalogue which lie within the error-box
of the 1.8 \MeV\ emission feature from which 14 have age estimates below
20 Myr, one falls in the age interval 20-100 Myr, 4 are older than 100 Myr
and 12 are without age estimates.
The remarkable correlation of the Carina feature with open clusters younger
than 20 Myr strongly suggests that WR stars and SNe in these clusters
are the origin of the observed 1.8 \MeV\ emission.

The Carina region is not only the region with the highest concentration
of young open clusters, it also houses one of the most prominent \HII\
regions of the Galaxy: the Carina nebula NGC 3372.
NGC 3372 is assumed to be powered by the four hot star clusters Tr 14,
Tr 15, Tr 16 and Cr 228 (\cite{rf:dmd86}).
These clusters comprise the largest concentration (six) of O3 stars
known in the Galaxy, three WR stars, and the remarkable luminous blue
variable (LBV) \etacar.
Together with some other young clusters, they form the Car OB1
association whose photospectrometric distance has been estimated to be
$2.7\pm0.2$ kpc (\cite{rf:tghh80}).
Thus, if the observed 1.8 \MeV\ emission comes from this association,
the measured flux implies a total \al26\ mass of $0.021\pm0.006$~\Msol\
in that region.

The presence of evolved stars like \etacar\ and the WR stars in Car OB1
indicates that \al26\ was ejected in the recent past to the ISM.
For \etacar, Langer \etal\ (1995) estimated a total \al26\ ejection of
$10^{-3}$ \Msol\ which excludes it as the only source responsible for
the observed emission.
Also the three WR stars with typical yields of $2\times10^{-4}$~\Msol\
fail to explain the measured flux.
The large number of O3 stars, however, suggests that very massive stars
are not exceptional in Car OB1 and other stars possibly experienced
similar stages in the recent past.
Some stars may already have finished their life and ejected \al26\ in
a supernova explosion.
Though under debate, there are indications of recent supernova activity
in the Carina nebula from observations of large scale motions
(\cite{rf:mlk84}), non-thermal radio emission (\cite{rf:tsk91}), and
X-ray emission (\cite{rf:chu93}).
We presently exploit a full stellar evolution model for the clusters in
Carina to estimate the total \al26\ input in the recent past
(\cite{rf:knoedl95b}).

However, it remains questionable whether the young open clusters of the
Car OB1 association can explain all of the measured 1.8 \MeV\ flux, which
would require about 20 LBVs like \etacar, 100 WR stars, or 200 SNe in the
last $10^6$ yrs, respectively.
There may be additional similar regions rich in massive stars along the
line of sight which contribute to the observed emission.
This possibility is suggested from the fact, that a Galactic spiral arm is seen
tangentially in the direction of Carina (e.g. \cite{rf:grabelsky87}).
\HII\ regions together with their exciting young open clusters are
generally associated with the spiral structure (\cite{rf:elmegreen85}).
In this context it should be mentioned that 3 of the 5 known Galactic
LBVs (\cite{rf:hd94}) are found in Carina, being all consistent with the
observed emission feature.
This indicates that evolved extremely massive stars are indeed present
behind Car OB1 whose ejected \al26\ should contribute to the observed
1.8 \MeV\ emission.

\section{Conclusion}

Prominent 1.8 \MeV\ \gray\ line emission which is attributed to the
radioactive decay of \al26\ was detected from the Carina region by the
imaging telescope COMPTEL.
We find a remarkable correlation of the 1.8 \MeV\ feature with a
concentration of open
clusters younger than 20 Myr which suggests that WR stars and core
collapse SNe are the origin of the observed emission.
The presence of evolved objects in the Carina clusters clearly indicates
recent \al26\ ejection to the ISM.
This region demonstrates that OB-associations and young open clusters could be
a natural explanation for some of the emission features and `hot-spots'
in the COMPTEL 1.8 \MeV\ sky-map.
Especially noncoeval star-formation can produce veritable starbursts
(\cite{rf:ss95}) which would result in substantial concentrations
of \al26\ within a localized region.
If such a burst occurred within the last million years it could be observable
through its strongly peaked 1.8 \MeV\ emission.
We will investigate further regions of peaked 1.8 \MeV\ emission to
verify this hypothesis.

\begin{acknowledgements}
The COMPTEL project is supported by the German government through
DARA grant 50 QV 90968, by NASA under contract NAS5-26645, and by
the Netherlands Organisation for Scientific Research NWO.
\end{acknowledgements}


\end{document}